\documentclass[aps,showpacs,floatfix,twocolumn]{revtex4}
\usepackage{graphicx}
\newcommand{\be}{\begin{eqnarray}}
\newcommand{\ee}{\end{eqnarray}}
\def\beq{\begin{equation}}
\def\eeq{\end{equation}}

\begin{document}
\title{Differentiable potentials and metallic states in disordered {\bf one-dimensional} systems}
\author{Antonio M. Garc\'{\i}a-Garc\'{\i}a}
\affiliation{Physics Department, Princeton University, Princeton,
New Jersey 08544, USA}
\affiliation{The Abdus Salam International Centre for Theoretical
Physics, P.O.B. 586, 34100 Trieste, Italy}
\author{Emilio Cuevas}
\affiliation{Departmento de F\'{\i}sica, Universidad de Murcia,
E-30071 Murcia, Spain}
\begin{abstract}
We provide evidence that as a general rule Anderson localization effects become weaker as the degree of differentiability of the disordered potential increases.
In one-dimension a band of metallic states exists provided that the disordered
potential is sufficiently correlated and has some minimum degree of
differentiability. Several examples are studied in detail. In agreement with the one parameter
scaling theory the motion in the metallic region is ballistic if the
spectral density is smooth. Finally, we study the most promising
settings to observe these results in the context of cold atoms.
 \end{abstract}

\maketitle
Eigenstates of a one-dimensional (1d) system are exponentially localized for any disorder and energy \cite{anderson,mott} provided that hopping is restricted to nearest neighbors and the potential is random and uncorrelated.
A natural question to ask is to what extent
this result still holds if these conditions are relaxed. The effect of long range hopping is well understood \cite{cuevas}.
Localization, though not necessarily exponential, persists in case that the
hopping term decays asymptotically as $1/n^\kappa$ with $\kappa > 1$ or faster.\\ 
By contrast, the effect of a correlated disorder
\cite{pastur,damanik1,damanik,kotani,kosi1,efetov,shomerus,russ,dun}
is still far from being completely settled. The recent realization
of disordered systems by using ultra cold atoms
\cite{aspect,fallani} in optical lattices has increased enormously
the interest in this problem since the experimental potential is
random but highly correlated.
We first review previous literature on localization in correlated potentials. According to
Kotani's theory \cite{pastur,damanik1,damanik,kotani,kosi1} of
random ergodic operators a necessary condition for the existence of
a metallic band is
that the potential be deterministic. A potential is deterministic if
given its behavior in a certain small interval it is possible to
predict its value in the rest of points \cite{pastur}. A Gaussian
potential $V(n)$ with correlation function $B(n) \equiv  \langle
V(n)V(0) \rangle$ is said to be non deterministic if and only if
\cite{bookran} $\int_{-1}^{1}dk {\log S(k)} > -\infty$, where $S(k)$
is the Fourier transform of $B(n)$. An important consequence of Kotani's theory \cite{kosi1}
is that a metallic band cannot exist if the potential is discontinuous \cite{damanik} or it is
a Gaussian noise with correlations such that $\lim_{n \to \infty} B(n) \to 0$ as power law or faster \cite{kosi1}.\\
In the physics literature
 \cite{izrailev} it was claimed that a band of metallic states exists in 1d 
 provided that $S(k) = 0$ in a certain range of $k's$.
 Some potentials with this property (for instance Gaussian disorder with $B(n) \propto \sin(bn)/n$ and $b>0$) are investigated in detail in Ref. \cite{izrailev}.
The approach of Ref. \cite{izrailev} uses the fact that, in
the Born approximation, the Lyapunov exponent is proportional to
$S(k)$. However we note that a) a vanishing Lyapunov exponent is not
a signature of a metallic state. It only shows that the decay of
eigenstates is slower than exponential,  b) no metallic band can
exit for $B(n) \propto \sin(bn)/n$ since $\lim_{n \to \infty} B(n)\propto 1/n \to
0$ as a power-law \cite{kosi1}. This is also 
confirmed by higher order perturbative calculations \cite{tessi}.\\
In Ref. \cite{lyra} it was conjectured that metallic states were
related to disordered potentials such that $S(k) \propto 1/k^\gamma$
with $\gamma > 2$. We note however that for  $B(n) \sim e^{- b
|n|^{c +1}}$ with $0 < b \ll 1$ and $0 \leq c < 1$,
$S(k) \propto 1/k^{c+2}$ for almost all $k$'s. However a
metallic band cannot exist \cite{kotani,kosi1}
since $\lim_{n \to \infty} B(n) = 0$.\\
These results place very strict but not insurmountable conditions on
the type of potentials that can lead to metallic states. A
paradigmatic exception is the case of quasiperiodic potentials.  For
$V(n) = \lambda \cos(2\pi \omega n +\theta)$ where $\omega$ is an
irrational number, $\theta \in [0,1]$, all eigenstates are
delocalized for $\lambda < 2$ \cite{jito0}.\\
The one parameter scaling theory (OPT) \cite{one} is a useful tool
to study localization effects. A key concept in the OPT is the
dimensionless conductance $g(N) = E_c/\Delta$ \cite{thouless2}
where, $E_c$ is the Thouless energy, $\Delta$ is the mean level
spacing and $N$, in 1d, it is system size. An insulator is characterized by $\lim_{N \to \infty} g(N) = 0$. 
The mean level spacing $\Delta \propto 1/N$ so in order to observe metallic behavior in 1d, $E_c \propto 1/N$.  
The typical time 
to cross the sample $t_c$ is related to the $E_c$ through the Heisenberg relation $t_c E_c \sim \hbar$.
The scaling of $E_c \propto 1/N$ corresponds thus to ballistic motion $t_c \propto N$.  
This is consistent with the results of \cite{simon5} where it was shown that quantum motion is slower than ballistic if eigenstates are exponentially localized.\\
A natural question to ask is whether it is possible to characterize which potentials lead to a 
band of metallic states in the associated Hamiltonian. The main goal of this paper is to answer affirmatively this question. 
We put forward a general relation between the degree of differentiability
of the potential and the magnitude of Anderson localization effects. We show numerically that potentials with
some minimum degree of differentiability and sufficiently strong long range
correlations produce a band of metallic states characterized by
quantum ballistic motion. For
quasiperiodic potentials we show analytically that metallic states exist provided $V(x) \in C^\beta$ with $\beta > 0$ where $C^\beta$ stands for functions
which are continuous and $\beta$--differentiable.
For non
quasiperiodic potentials metallic states exist provided that the continuous limit of $V(n) \in C^{\beta}$ with $\beta
> 1/2$. 
There are several reasons that indicate that eigenstates localization and differentiability of the potential are related: in 1d systems with uncorrelated
disorder eigenstates are always exponentially localized. Localization effects are so strong in 1d because the transmission and reflection probability for different sites are completely uncorrelated quantities. As a consequence the total probability of reflection never vanishes and eventually the particle gets localized. By contrast a certain degree of differentiability assures the potential in neighboring sites is strongly correlated. It is thus plausible that for sufficiently differentiable potentials a band of metallic states occurs due to destructive
  interference effects in the reflected components of the wavepacket. This is similar to the mechanism of delocalization in 1d random dimer models \cite{dun}.
Differences in the minimum degree of differentiability are expected to depend on whether the potential is quasiperiodic or not.
In the former localization can be avoided either by resonant tunneling or by enhanced destructive interference due to the smoothness of the potential. In the latter only the second mechanism is at work.\\
{\it Results.-}
We combine analytical techniques, a finite size scaling analysis \cite{sko} and a detailed study of $g(N)$ in order to explore the existence of metallic states in 1d systems.
To carry out the finite size scaling analysis we compute eigenvalues of the different Hamiltonians of interest by using standard numerical diagonalization techniques. For a given disorder and energy window the number of eigenvalues
obtained is at least $2 \times 10^5$.
The dimensionless conductance (transmission) is calculated by using the transfer matrix
method (see [\onlinecite{markos06}] and references therein). Fluctuations were reduced by computing $\langle \ln g(N) \rangle$
where, for a given energy, $\langle \ldots \rangle$ stands for ensemble average over at least $10^5$ disorder realizations.\\
The finite size scaling method \cite{sko} is based on the study of
the scaling properties of a spectral correlator. A popular choice is
the variance $\rm var(s)$ of the level spacing distribution $P(s)$,
where $P(s)$ is the probability of finding two neighboring
eigenvalues at a distance $s = (\lambda_{i+1} - \lambda_{i})/\Delta
$ and \be \label{var} {\rm var(s)} \equiv \langle s^2 \rangle -
\langle s \rangle^2
  = \int_0^{\infty}ds ~s^2 P(s)- 1,
\ee
where $\langle \dots \rangle$ denotes spectral and ensemble averaging.
The prediction for a metal with time reversal invariance is
${\rm var(s)} \approx 0.273$  (${\rm var(s) = 0}$) if the motion is diffusive
(ballistic) while for an insulator gives ${\rm var(s)} = 1$.
If the variance gets closer to the metal (insulator) result as the
 volume is increased we say that the system is a metal (insulator).\\
{\it Quasiperiodic potentials.-}
In this section we explore the relation between differentiability and 
localization in quasiperiodic potentials. As was mentioned previously metallic states do exist for analytical 
potentials $V(n) \propto \cos(\omega n +\theta)$ \cite{jito}. In \cite{jito0} it was proved the existence of metallic states in
less smooth potentials
 $V(n) = \sum_k a_k \cos(\omega n +\theta)$ with $|a_k|< Ae^{-Bk}$, $A,B$ positive constants, and $\omega$ an irrational number.
It is conjectured \cite{last} a metallic band might exist for 
$V(n) \in C^{\beta}$ and $\beta > 3/2$.
Below we provide evidence of the existence of metallic states 
for even less smooth potentials $V(n) \in C^\beta$ with $\beta > 0$.\\
Our starting point is a 1d tight binding Hamiltonian, \be
\label{main} {\cal H} \psi_n = \psi_{n+1} + \psi_{n-1} +
\frac{1}{\lambda}V(\omega n +\theta)\psi_n\,, \ee where $V(x) =
\sum_k a_k \cos(2\pi k x)$, $\theta \in [0,1]$ and $a_k$ are real
coefficients. From this definition is clear the $V(x) \in
C^{\beta}$ provided that $|a_k| < A/k^{1+\beta}$ with
$\beta$ and $A$ real positive constants. This model is in
principle suitable to an analytical treatment.
The first step is to Fourier transform Eq.(\ref{main}):
\be \label{main1} {H} \psi_k = \sum_m a_{k-m} {\psi}_{m} +
2\lambda\cos(2\pi\omega k +\theta)\psi_k \,, \ee where, 
$H$ is, after a $\lambda$ rescaling, the Fourier trasform of $\cal H$. 
We note
that
according to Ref.\cite{cuevas} all eigenstates of
Eq.(\ref{main1}) are localized for $\beta > 0$ provided the
diagonal element is random uncorrelated instead of $\sim
\lambda \cos(2 \pi \omega k + \theta)$. However for sufficiently large
$\lambda$ this potential leads to a full band of localized states
\cite{jito}. It is thus plausible that its pseudo random character
is not important in this limit and consequently the results of
Ref.\cite{cuevas} apply. Then it remains to show that localization
in the Hamiltonian Eq.(\ref{main1}) means delocalization for the
Hamiltonian Eq.(\ref{main}). This fact was proved  in
Ref.\cite{gordon} for quasiperiodic potentials such as the one of
Eq.(\ref{main}). Numerical results, not shown, fully confirm this
picture. In conclusion, a band of metallic states can exist provided
that $ V(x) \in C^\beta$ with $\beta >0$.\\
\begin{figure}
\includegraphics[width=0.99\columnwidth,clip]{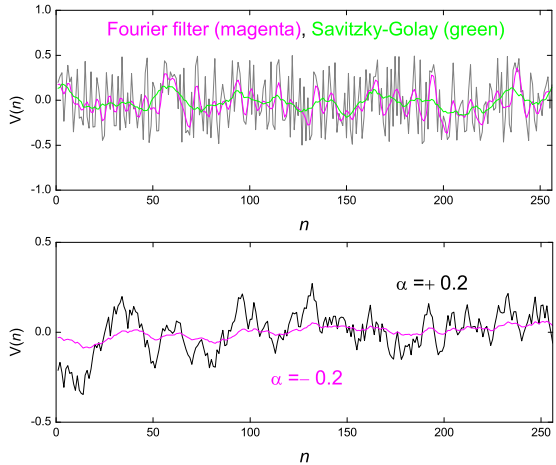}
\caption{(color online) Upper: $V(n)$ after smoothing an uncorrelated random potential, grey line, by a Fourier filter, magenta line (dark grey), and the Savitzky-Golay method, green line (light grey).
 Lower: Eq. (\ref{our}) for $\beta = \alpha + 1/2 = 0.7$ black line and $\beta = \alpha +1/2 = 0.3$ pink line (light grey).}
\label{fig1}
\end{figure}
\begin{figure}
\includegraphics[width=4.2cm,height=0.55\columnwidth,clip]{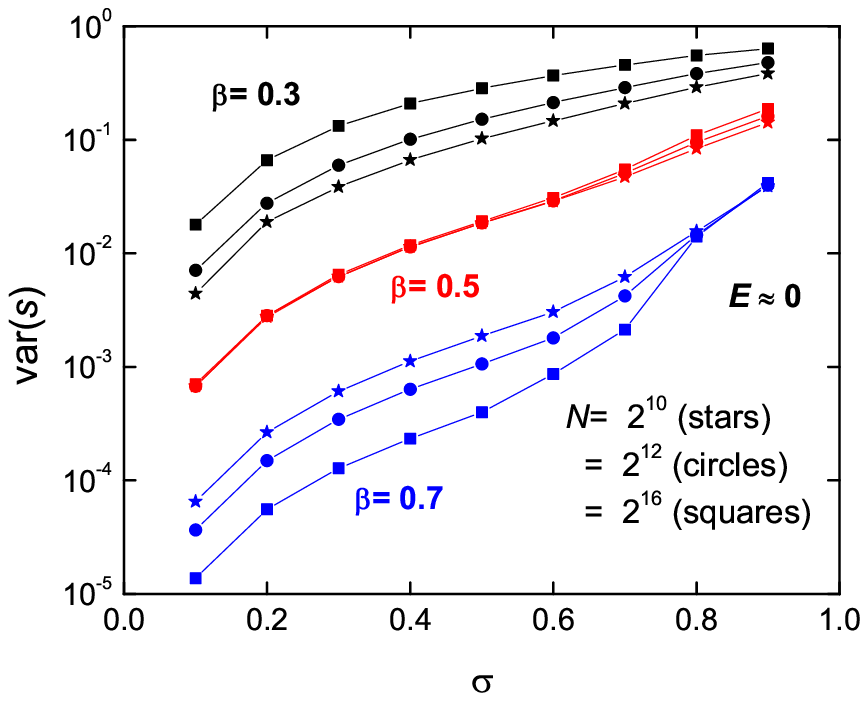}
\includegraphics[width=4.2cm,height=0.55\columnwidth,clip]{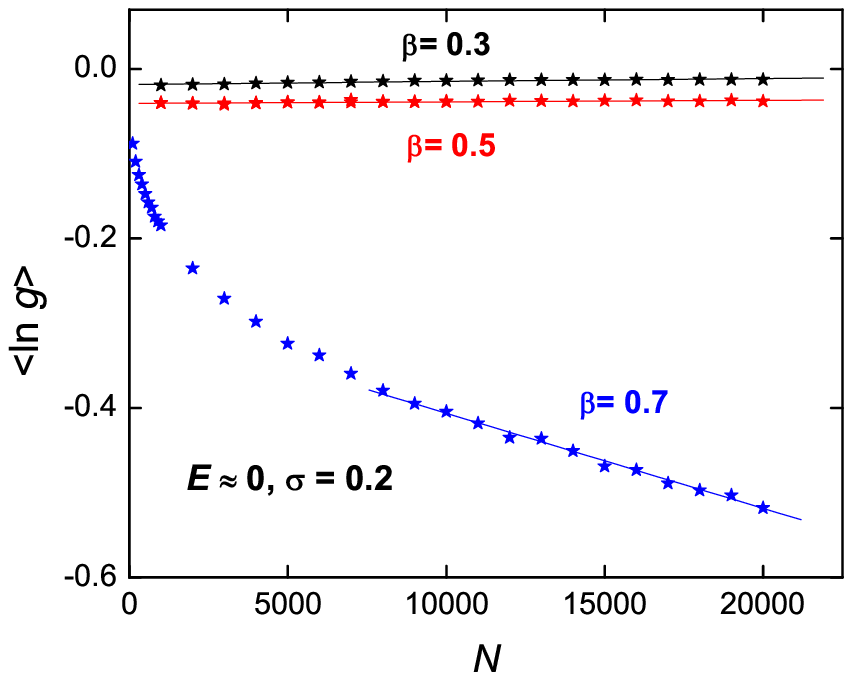}
\caption{(color online) Left: ${\rm var(s)}$, Eq. (\ref{var}), as a function of disorder for $E \approx 0$ and different sizes $N$
for a system with $V(n)$ given
by Eq. (\ref{our}).
Metallic states are only observed for $V(n)\in C^{\beta}$ with $\beta > 1/2$.
Right: dimensionless conductance $g(N)$
as a function of the system size $N$, $\sigma = 0.2$, $E \approx 0$ and $\beta = 0.7,0.5,0.3$. In agreement with the $\rm var(s)$ results, metallic states, characterized by a constant $g$, only occur for $\beta > 1/2$.}
\label{fig2}
\end{figure}
{\it Non quasiperiodic systems.-}
In this section we show numerically that for non quasiperiodic potentials a band of metallic states 
can only occur for $V(n) \in C^{\beta}$ with $\beta > 1/2$. Our findings, though obviously consistent 
with the Kotani's theory \cite{kotani} mentioned in the introduction, cannot be obtained from it.
We note this theory only provides necessary ($\lim_{x \to \infty} B(x) \neq 0$) but no sufficient conditions for the   
existence of metallic states. \\
In order to 
generate a potential with a given degree of differentiability we smooth a 
Gaussian uncorrelated potential by using different methods available in the literature:  Savitzky-Golay \cite{golay}, Fourier filter and 
fractional integration method \cite{fraci,grun}. Results should not depend on the smoothing method provided 
that both 
the degree of differentiability and the correlations of the resulting smoothed potential are the same. 
Our first smoothing method consists in the application of a fractional integral operator, the Gr\"{u}nwald-Letnikov operator \cite{grun}, on an uncorrelated random potential. 
The smoothed potential is given by, \be
\label{our} V(n) = D^{(-\beta -1/2)} a_n =\sum_{i=0}^{n} (-1)^i
{{-\beta -1/2}\choose i}a_{n-i} \,, \ee where $D^{(-\beta -1/2)}
a_n$ stands for the $\beta + 1/2$--integral of the random potential $a_n$. According to Kotani's theory a necessary 
condition for the existence of metallic states is that $V(n)$ be deterministic \cite{kosi1,pastur}.
This can be achieved by choosing $a_n$ from a Gaussian
distribution with a $N$ dependent variance $\propto 1/N^{\beta}$.
Finally we carry out a $N$ independent rescaling of the potential
such that $\langle V(n)\rangle = 0$ and $\langle V^2(n) \rangle =
\sigma^2$ (see Fig. 1). With these definitions: a) for $\beta \geq 0$ the continuous limit
of the potential belongs to 
$C^{\beta}$, b) for
$\beta
< 0$ all states are localized for any $\sigma$ since the potential is
discontinuous \cite{kotani,kosi1,damanik}, c) the correlations are
such that $B(n)$ is bounded and $\lim_{n \to N} B(n) \neq 0$ for $N
\to \infty$. This is consistent with the Kotanis's result \cite{kosi1} that no metallic states can if $B(n)$ decays as a power law or faster for large $n$. \\
 We carry out a finite size scaling analysis of the spectrum
combined with a study of $g(N)$ for different values of
$\beta$. 
In Fig. 2 (left) we plot $\rm var(s)$ around the center of the band
$E \approx 0$ as a function of $\sigma$ for different $\beta's$. It is clearly  
observed that for $\beta < (>) 1/2$ the variance $\rm var(s)$ increases (decreases) with the system size. This is a 
signature of an insulator (metal). For $\beta = 1/2$, $\rm var(s)$ is almost scale invariant for sufficiently weak disorder.
Larger volumes would be needed to determine its localization properties.
The behavior of the dimensionless conductance further agrees with this picture, Fig. 2 (right). In agreement with the OPT, the dimensionless conductance is size independent around $E \approx 0$ for any $\beta \geq 1/2$ and weak disorder.
In conclusion, for sufficiently
weak disorder we found a band of metallic states for $V(n) \in C^{\beta}$ and $\beta > 1/2$.
Thus metallic states exist provided that $V(n)$ is at least $1/2$--differentiable and correlations are strong enough $\lim_{n \to \infty}B(n) \neq 0$.\\
We further test the relation between differentiability of the potential and Anderson localization by studying a 1d system with an
uncorrelated random potential which is subsequently smoothed either
by the Savitzky-Golay \cite{golay} or the Fourier filter method (see
Fig. 1).\\
The Savitky-Golay method permits to smooth an initial uncorrelated potential by the best fit of a polynomial of degree $M$ of the
$n_p$ surrounding a given point of the original uncorrelated
potential. The potential is correlated only up to distances $~n_p$ therefore $\lim_{n \to \infty} B(n) = 0$. 
According to Kotani's theory \cite{damanik,kosi1} metallic states can only exist if $\lim_{n \to \infty} B(n) \neq 0$.\\
In the Fourier filter method a smoothed potential is obtained following three steps: a) the uncorrelated potential is 
Fourier transformed, b) the transformed data is processed in the
$k-$domain using the window function $w(k)=1 -(k/k_c)^2$ with $k_c
= N/n_p$ a given cut-off, c) the
modified signal is transformed back to real space. In this case the
resulting potential is clearly analytical and $\lim_{n \to \infty} B(n) \neq 0$ so a band of metallic states might occur
for sufficiently weak disorder.\\
 A finite size scaling analysis (see Fig. 3) fully confirms that only the Fourier filter
method leads to a metallic band for $E < E_c \approx 1$ characterized again by ballistic
motion ($\rm var(s) = 0$).
By contrast no metallic states are observed if the smoothed potential is obtained by 
the Savitky-Golay method. \\
\begin{figure}
\includegraphics[width=4.2cm,height=0.55\columnwidth,clip]{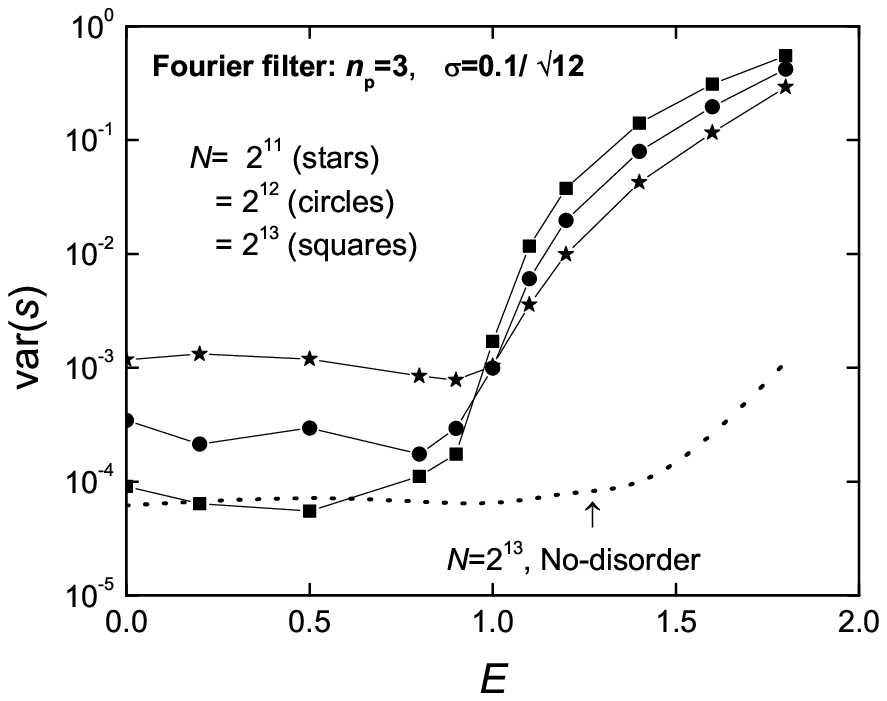}
\includegraphics[width=4.2cm,height=0.55\columnwidth,clip]{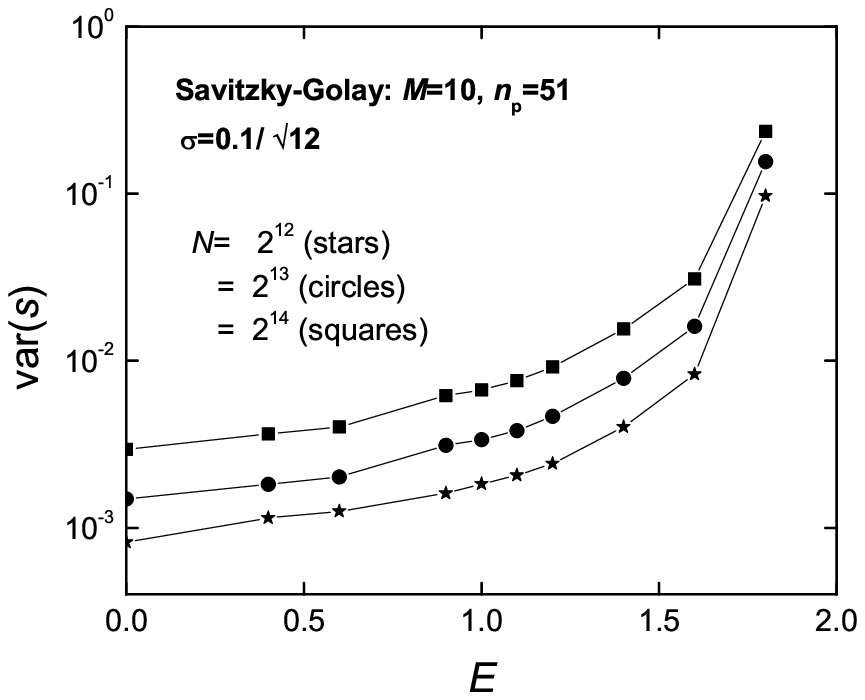}
\caption{${\rm var(s)}$ versus energy for different system sizes.
 $V(n)$ is a random uncorrelated potential of zero mean
and $\sigma = 0.1/\sqrt{12}$ smoothed by (Left) a Fourier filter with $n_p=3$
and (Right) the Savitzky-Golay filter with $M=10$ and $n_p=51$ (see text for details). The dotted line in
the left panel corresponds to the ${\rm var(s)}$ of a periodic sample. Metallic states only exist in the case of the Fourier filter method.}
\label{fig3}
\end{figure}
{\it Experiments with cold atoms.-}
We explore different possibilities to test experimentally the results of previous sections by using cold atoms in
 speckle potentials \cite{aspect} and multichromatic lattices \cite{fallani}.
A speckle pattern is formed by diffraction of a laser beam through a
rough plate \cite{aspect}. The resulting speckle potential felt by
the cold atom is random but correlated. A typical signature of these
potentials is that $S(k)$ vanishes for $|k| > k_c$ where $k_c$
depends on the details of potential. For instance in
Ref.\cite{aspect}, $B(n)  \sim \frac{\sin^2(n/\delta)}{n^2}$ where
$\delta$ is the speckle grain size and $k_c \sim 1/\delta$. From the
mathematical results reviewed in the introduction it is clear that
these correlations will increase the correlation length \cite{tessi}
but will not induce an Anderson transition
\cite{kotani,kosi1}.\\
A more promising option is to consider the random non ergodic
potential $V(n) = \xi_n/n^\kappa$,
where $\xi_n$ are random number from a box or Gaussian distribution  \cite{sou}. Quantum dynamics depends strongly on the value of $\kappa$ \cite{sou}.
For $\kappa < 1/2$ all states are localized. For $\kappa \geq 1/2$ there is a metal-insulator transition in a certain region of energies or disorder.
These results are not expected to be modified if $V(n)$ is weakly correlated ($\lim_{n \to \infty} B(n) \to 0$).
A speckle pattern such that the resulting potential has a decreasing intensity is within the reach of current
technical capabilities.\\
A multichromatic lattice \cite{fallani} is created by combining several standing light waves with different, non commensurate frequencies.
In this case the potential is not random but rather quasiperiodic.
It is thus not surprising that for weak disorder a region of
metallic states with ballistic motion was observed \cite{fallani}.
The potential resulting after a Fourier filter smoothing studied
above could in principle be modeled with these techniques.\\
To conclude we have put forward a general characterization for the existence of metallic
states in 1d systems. Our main results are: a) the degree of differentiability of the potential 
acts as a control parameter to induce a metal insulator transition, 
b) a metallic band exists in 1d provided $\lim_{n \to \infty}
B(n) \neq 0$ and $V(n) \in C^{\beta}$ with $\beta > 0 (1/2)$ for (non) quasiperiodic potentials, c) cold atoms techniques might be suited to observe
the metal-insulator transition in 1d, d) in agreement with OPT, the quantum dynamics is
ballistic for a 1d metal.

We are indebted to G. Shlyapnikov and S. Jitomirskaya for a
critical reading of the manuscript. We thank thank M. Aizenman, S.
Warzel, S. Kotani, H. Shomerus and M. Caceres for fruitful
conversations. AMG acknowledges financial support from a Marie Curie
Outgoing Action, contract MOIF-CT-2005-007300. We thank the FEDER
and the Spanish DGI for financial support through Project No.
FIS2007-62238.


\end{document}